\documentclass{article}
\usepackage{amssymb}
\usepackage{bbding}
\usepackage{pifont}
\usepackage{utfsym}
\usepackage{fontawesome}
\usepackage{dcase2022,amsmath,graphicx,url,times,booktabs, tabularx}

\title{Segment-Level Metric Learning for Few-shot Bioacoustic Event Detection}

\name{
      Haohe Liu$^{1}$,
      Xubo Liu$^{1}$,
      Xinhao Mei$^{1}$, 
      Qiuqiang Kong$^{2}$,
      Wenwu Wang$^{1}$,
      Mark D. Plumbley$^{1}$
}
\address{$^1$ Centre for Vision, Speech, and Signal Processing~(CVSSP), University of Surrey, UK\\ 
        $^2$ Speech, Audio, and Music Intelligence (SAMI) Group, ByteDance, China \\
}

\begin{document}

\ninept
\maketitle

\begin{sloppy}

\begin{abstract}
Few-shot bioacoustic event detection is a task that detects the occurrence time of a novel sound given a few examples. Previous methods employ metric learning to build a latent space with the labeled part of different sound classes, also known as positive events. In this study, we propose a segment-level few-shot learning framework that utilizes both the positive and negative events during model optimization. Training with negative events, which are larger in volume than positive events, can increase the generalization ability of the model. In addition, we use transductive inference on the validation set during training for better adaptation to novel classes. We conduct ablation studies on our proposed method with different setups on input features, training data, and hyper-parameters. Our final system achieves an F-measure of 62.73 on the DCASE 2022 challenge task 5~(DCASE2022-T5) validation set, outperforming the performance of the baseline prototypical network 34.02 by a large margin. Using the proposed method, our submitted system ranks 2nd in DCASE2022-T5. The code of this paper is fully open-sourced\footnote{\url{https://github.com/haoheliu/DCASE_2022_Task_5}}.
\end{abstract}

\begin{keywords}
few-shot learning, transductive inference, metric learning, audio event detection
\end{keywords}

\section{Introduction}
\label{sec:intro}
Few-shot learning~(FSL)~\cite{ravi2016optimization} is a machine learning problem that makes predictions based on the training data that contains limited information.
Sound event detection~(SED)~\cite{mesaros2021sound} is a task that locates the onset and offset of certain sound classes. By combining the idea of FSL with SED~\cite{wang2020few}, a system can detect a new type of sound with only a few examples. Few-shot SED is useful for audio data labeling, especially when the user needs to detect a new type of sound.

Most prior studies use a prototypical network~\cite{snell2017prototypical} as the main architecture ~\cite{yang2021few, tang2021two, zhang2021few, anderson2021bioacoustic}. Yang et al.~\cite{yang2021few} propose a mutual learning framework that employs transductive inference to iteratively improve the feature extractor and classifier, where transductive inference means the model has access to the test set without labels during the training process. A smoother manifold of embedding space can help extend the decision boundary and reduce the noise in data representation~\cite{rodriguez2020embedding}. Tang et al.~\cite{tang2021two} propose to use embedding propagation~\cite{rodriguez2020embedding} in few-shot SED to learn a smoother manifold by interpolating between the model output features based on a similarity graph.
Data augmentations such as spec-augment and mixup are used in the method described in~\cite{zhang2021few, anderson2021bioacoustic}. 
There is also a spectrogram-cross-correlation-based method called template matching~\cite{morfi2021few}, which performs detection based on the normalized cross-correlation between example sound event and unlabeled data. 


Metric learning~\cite{kulis2013metric} refers to learning a distance function and feature space for a task. Previous metric-learning-based studies~\cite{yang2021few, tang2021two} usually optimize the model with the labeled positive events, by grouping and separating the latent prototypes of the events with the same and different classes, respectively. The audio chunks that do not contain target events, which we refer to as negative events, are larger in volume but receive less attention. For example, in the DCASE 2022 task 5 development set~\cite{morfi2021few}, the duration of the negative events is 19.18 hours, accounting for 91.3\% of the training data with a total duration of 21 hours. 




\begin{figure}[tbp]
    \centering
    \includegraphics[page=3, width=0.9\columnwidth]{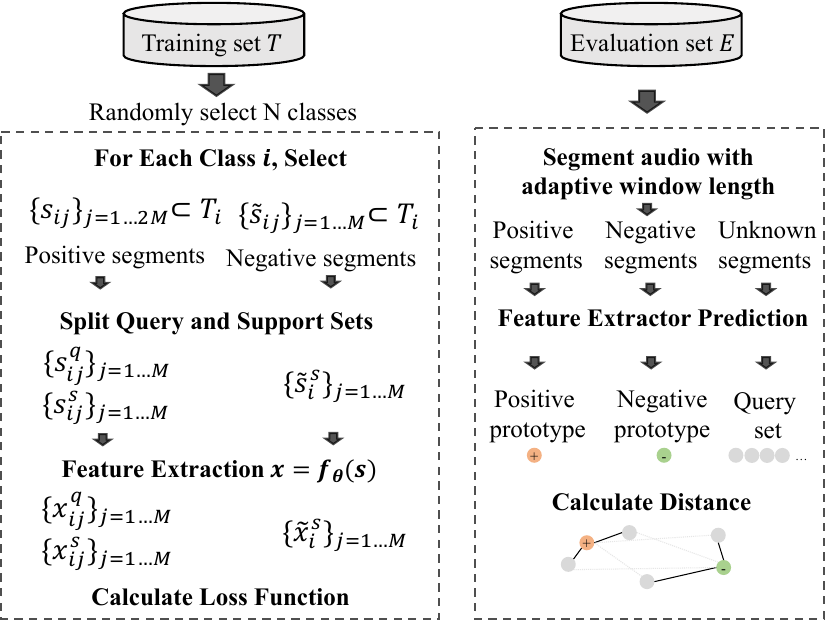}
    \caption{Training and evaluation procedure of the N-way-M-shot segment-level metric learning. $M$ denotes the number of segments or embeddings.}
    \label{fig:system-overview}
    \vspace{-5mm}
\end{figure}



In this paper, we propose a segment-level metric learning method that achieves state-of-the-art results on the few-shot bio-acoustic detection task. As shown in Figure~\ref{fig:system-overview}, our system operates on a segment level. Each sound event can contain multiple segments. We train a feature extraction network that maps the segments into latent embeddings, which are averaged into prototypes to represent different sound classes.
To learn a robust latent space, we use a transductive learning scheme and propose to build the contrastive loss with negative events. We also improve our method by using feature selection, data augmentation, and post-processing.
We perform ablation studies to measure the effectiveness of each component. Our proposed method achieves an \text{F-measure} of 62.73 on the DCASE task 5 validation set. 


This paper will be organized as follows. Section~\ref{sec:overview} provides an overview of our system. Section~\ref{sec:methodology} introduces our methodology. Section~\ref{sec:experiments} discusses the experimental setup. Section~\ref{sec:result} reports the evaluation result and the ablation studies. Section~\ref{sec:conclusions} summarizes this work and provide a conclusion.


\section{System Overview}
\label{sec:overview}

\begin{figure}[tbp]
    \centering
    \label{fig:negative-enhanced-contrastive-learning}
    \vspace{-1mm}
    \includegraphics[page=1, width=\columnwidth]{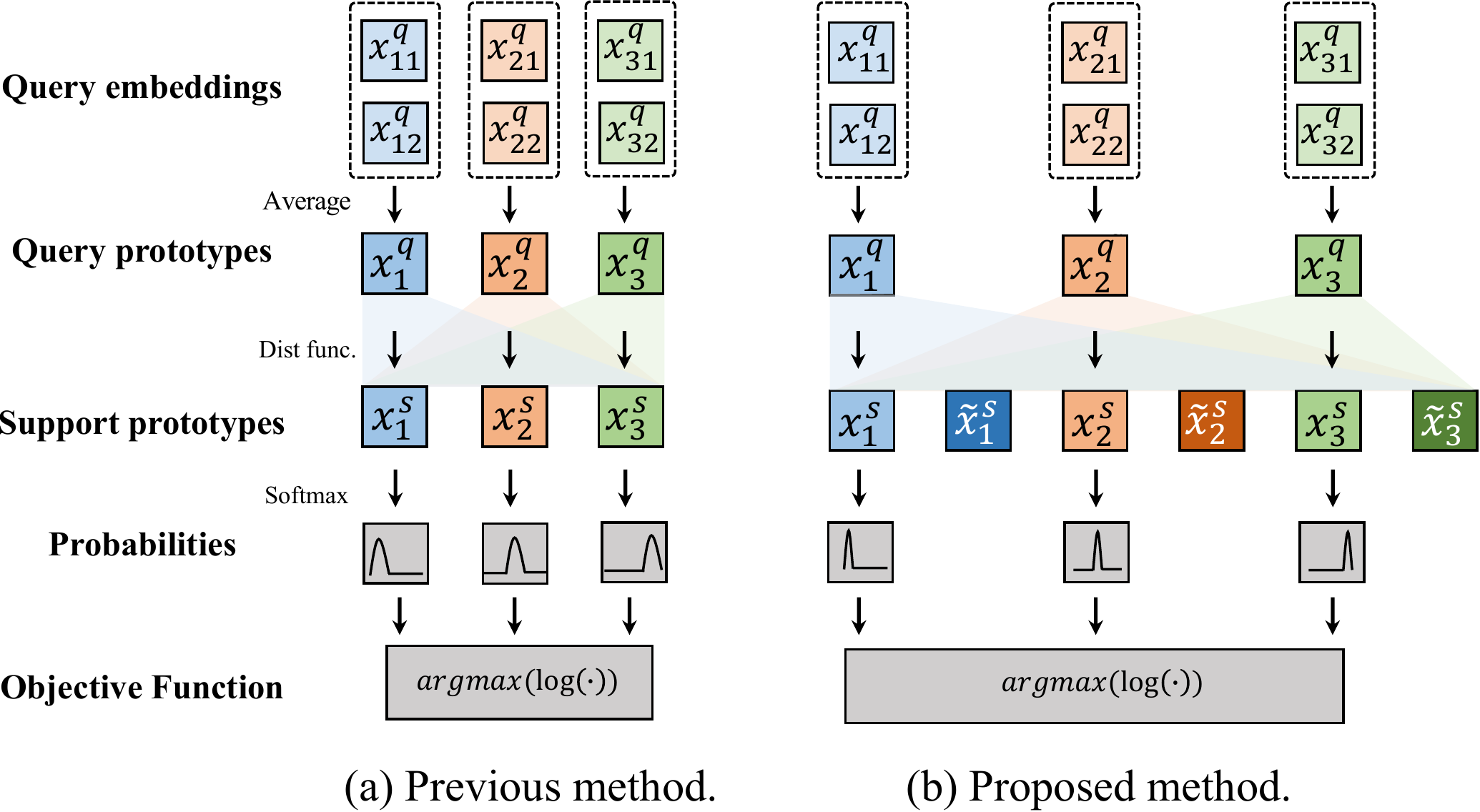}
    \caption{This figure illustrates $N$-way-$M$-shot metric learning when $N$=3 and $M$=2. (a) is a visualization of previous method, which only uses positive classes. (b) shows the proposed metric learning with the negative segments we used in our system. Support embeddings are omitted for simplicity. $x_{i}$ and $\widetilde{x}_{i}$ stand for the positive and negative prototypes of class $i$.}
    \vspace{-4mm}
\end{figure}

We build our system using a prototypical network~\cite{snell2017prototypical}, which is widely used for metric-based few-shot learning. The training data $T=(S_{i}, Y_{i})|_{i=1}^{N_{train}}$ contains audio feature set $S_i=\{s_i|y_i=1\} \sqcup \{\widetilde{s}_i|y_i=0\}$ and its corresponding label set $Y_i=\{y_i| y_i\in\{0,1\}\}$, where $\{s_i\}$ and $\{\widetilde{s}_i\}$ are the sets of positive and negative segments for class $i$, respectively, and $N_{train}$ is the total number of training segments.
The evaluation dataset $E=(S^{\prime}_{i}, Y^{\prime}_{i})|_{i=1}^{N_{eval}}$ also contains an audio feature set $S^{\prime}_{i}=\{s^{\prime}_{i}\}$ and a label set $Y^{\prime}_{i}=\{y^{\prime}_{i}\}$, where $N_{eval}$ is the number of classes in the evalution set, $|S^{\prime}_{i}|=L_{i}$ and $|Y^{\prime}_{i}|=K$. Here we have $L_{i} \ge K$ because the evaluation set is partially labeled with only first $K$ events.
The objective of our system is properly mapping different audio features into a latent embedding within a high-dimensional space, where similar audio features are closer together. 

We use episodic training~\cite{li2019episodic} to optimize our system in an N-way-M-shot way. As illustrated in Figure~\ref{fig:negative-enhanced-contrastive-learning}(a), N-way-M-shot means each training batch will select data from $N$ classes. And for each classes $i$, the system will randomly select $M$ segments $\{s^{s}_{ij}\}_{j=1...M}$ as support segments and another $M$ segments $\{s^{q}_{ij}\}_{j=1...M}$ as query segments. All the segments in different classes have the same length. Then a feature extraction network~(Section~\ref{sec:feature-extraction-network}) will map these segments into fix-length embeddings, which are later averaged into query prototypes $x^{q}_{i}$ and supporting prototypes $x^{s}_{i}$. The system is optimized by minimizing the distance between the query and support prototypes with the same class. 
To build a robust latent space and generalize better to the new class, we propose to use the negative event in metric learning and the transductive inference scheme in Section~\ref{sec:negative-metric-learning} and~\ref{sec:transductive-inference}. 

During evaluations, the audio file will be segmented using a sliding window with an adaptive segment length (Section~\ref{sec:ada-seg-len}). The segments in the labeled parts will be used to build positive and negative prototypes, which are treated as the latent representation of the positive and negative events in an audio file.
The segments in the unlabeled part are the query set, which can be classified by calculating and comparing the distance with the positive and negative prototype (Section~\ref{sec:negative-prototypes}). And if the probability of one query belonging to a positive prototype is greater than a threshold $h$, it will be classified as positive. Consecutive positive predictions will be merged into one single event.



\section{Methodology}
\label{sec:methodology}
\subsection{Feature extraction network}
\label{sec:feature-extraction-network}
Our feature extraction network $f_{\theta}$ is a convolutional neural network~(CNN) based architecture that maps the audio feature $s$ into a latent embedding $x$.
In a similar way to the architecture proposed by~\cite{kong2020panns}, the network $f_{\theta}$ consists of three convolutional blocks with hidden channels of sizes 64, 128, and 64. Each convolutional block consists of three two-dimensional CNN layers with batch normalization and leaky rectified linear unit activations~\cite{xu2015empirical}. As a common trick in CNN-based network~\cite{kong2020panns, liu2020channel}, we apply $2\times 2$ max-pooling after each block for downsampling and enlarging the reception field. The input and output of each convolutional block have a residual connection processed by a downsampling CNN layer. In order to maintain the same output dimension with different input lengths, we apply an adaptive average pooling at the end of the network. The final output feature map after adaptive pooling is a $C\times T\times F$ size block, which is the final latent embedding of $s$. 

\subsection{Segment-level metric learning}
\label{sec:negative-metric-learning}


We propose to utilize negative segments within negative events during model optimization to learn a more robust representation, as illustrated in Figure~\ref{fig:negative-enhanced-contrastive-learning}(b). In a similar way to ~\cite{wang2020few}, we first divide the audio features into segments with equal length for metric learning. Then $f_{\theta}$ maps all the segments into latent embeddings.
During optimization, we will calculate the class probabilities distributions of the query prototype $x_{i}^{q}$, which involves the distance calculation with all the positive and negative support prototypes. In this case, the model can learn a larger amount of contrastive information from the negative events on building the latent space. 
Specifically, we first calculate a distance matrix $\mathbf{D}=[\mathbf{d}^{(1)}, \mathbf{d}^{(2)}, ..., \mathbf{d}^{(N)}]^{T}$ according to Equation~\ref{eq:distance}, 

\vspace{-1mm}
\begin{equation}
    \label{eq:distance}
    \mathbf{d}^{(i)}_{2j} = \sqrt{\Sigma(x^{q}_{i}-x^{s}_{j})^{2}},~\mathbf{d}^{(i)}_{2j+1} = \sqrt{\Sigma(x^{q}_{i}-\widetilde{x}^{s}_{j})^{2}},
\end{equation}

where $\mathbf{d}^{(i)}\in \mathbb{R}^{2N}$ stands for the distance between $x^{q}_{i}$ and $2N$ support prototypes, and $\widetilde{x}^{s}_{j}$ denotes the support prototype for the negative events of class $j$. Then we optimize our model by maximizing the probability that $x^{q}_{i}$ is close to the positive support prototype of class $i$, $x^{s}_{i}$, 
given by




\vspace{-1mm}
\begin{equation}
    \mathbf{d}^{\prime(i)}=\textrm{log}(\textrm{Softmax}(\mathbf{d}^{(i)}))),~l = \textrm{argmax}_{\theta}(\Sigma_{i=1}^{N}(\mathbf{d}^{\prime(i)}_{2i}),
\end{equation}

where $0 \leq i,j \leq N, i,j \in \mathbb{N}$, and $l$ is the objective function. Note that the learning process does not involve the query prototypes for negative events $\widetilde{x}^{q}_{i}$, because $\widetilde{x}^{q}_{i}$ and $\widetilde{x}^{s}_{i}$ are not guaranteed to have the same type of sound. 

Data balancing is important in this task because different sound classes have different total durations~\cite{morfi2021few}. In order to balance between classes, we sample each class with equal probability during the episodic training. In this way, the model has equal probabilities to attend to each class and will be less prone to overfitting~\cite{gong2021psla}. 



\subsection{Transductive inference}
\label{sec:transductive-inference}

We adopt a transductive inference~\cite{boudiaf2021few} approach during training, which means our model will be optimized both on the fully-labeled training set and the partially labeled evaluation data. 
Each file in evaluation data has first $K$ labeled events for a particular type of sound. We treat these $K$ events as positive events and the remaining $K$ chunks of audio in the labeled part as negative events. 
In the evaluation set, although the sound class of each file is not available, files with same sound class should in the same subfolder, and we treat each subfolder of the evaluation set as a different sound class. Even though the files within each subfolder may not always contain the same target sound, our experiment shows transductive inference in this way can still help the model gain better adaptation to the evaluation set (Section~\ref{tab:ablation-study}). 





\subsection{Adaptive segment length}
\label{sec:ada-seg-len}

We use the same segment length among all classes during training for the convenience of batch processing. But during evaluation, using the same segment length is not ideal. For example, using a segment length that is too long or too short will tend to have a high false negative rate or false positive rate, respectively. In the evaluation set, different animal or bird species have drastically different lengths of vocalization, ranging from 30 milliseconds to 5 seconds. Thus we choose to use adaptive segment lengths during evaluation. 

\begin{table}[htbp]
\centering
\begin{tabular}{c|c|c|c|c|c}
\hline
$t_{\max}$ (s)  & {[}0,0.1{]} & (0.1,0.4{]} & (0.4,0.8{]} & (0.8,3.0{]} & (3.0,~$\infty$) \\ \hline
Length &    8    &  $t_{\max}$       &  $t_{\max}$ / 2         & $t_{\max}$ / 4          & $t_{\max}$ / 8        \\ \hline
\end{tabular}
\caption{The segment length we use on dividing the evaluation audio file for different values of $t_{\max}$.}
\label{tab:seg-len-setting}
\end{table}

As shown in Table~\ref{tab:seg-len-setting}, we set different segment length for each audio file based on the max length of the labeled events $t_{\max}=\max(t_{1},...,t_{K})$, where $t_{1},...,t_{K}$ denotes the duration of the $K$ labeled positive events. We set the hop length as one-third of the window length. 
Note the parameters here are chosen by experience.

\subsection{Positive and negative prototypes}
\label{sec:negative-prototypes}
During evaluation, we assume the first $K$ labeled positive events do not contain too much variety, therefore we calculate the positive prototype by averaging the embeddings of the labeled positive segments. By comparison, building negative prototypes is more tricky because negative segments can contain many different kinds of sounds. So simply averaging all the negative embeddings would result in a sub-optimal representation of negative prototypes.
To address these challenges, we choose to run our evaluation six times, each selecting 30 randomly selected negative segments within the labeled negative parts, and we average the predicted probabilities across time of six runs as the final prediction. The negative prototype in each run can have a chance to represent different sounds. This process is similar to the random subspace method~\cite{ho1998random}, in which the ensemble of several estimators trained with a different subset of training data can outperform a single estimator optimized on full training data.

\section{Experiments}
\label{sec:experiments}

\subsection{Dataset}


\textbf{DCASE2022-T5} The DCASE 2022 task 5 dataset\footnote{\url{https://zenodo.org/record/6482837}} contains a training set, a validation set, and an official evaluation set. The training and validation set are both fully labeled. The official evaluation set has the labels of the first five positive events. The full label of the official evaluation set is not released at the time of writing, so, we treat the validation set as the evaluation set in this paper following~\cite{Liu2022a, Du2022a}. The validation during training is not meant to pick the best model. That's because we perform validation in a different way from evaluation. 
Similar to the training process, we calculate validation accuracy on a fix-length segment level without adaptive segment length. 
Therefore the best model on validation does not necessarily perform the best during evaluation. Nevertheless, we use the same validation process in our experiments, so the comparisons are fair in different settings. There are also similar ideas in~\cite{kong2021decoupling, gong2021ast}, which utilize the evaluation set for validations. 

\label{sec:AudioSet-external_data}
\noindent
\textbf{AudioSet-Aminal-SL} AudioSet~\cite{gemmeke2017audio} is a large-scale dataset for audio research~\cite{kong2020panns, kong2021speech}. Considering that the training set of DCASE2022-T5 only contains 47 different sound classes, we choose to use the strongly labeled part of the AudioSet dataset\footnote{\url{https://research.google.com/audioset}} to augment training data with a wider variety of sounds. To alleviate the domain mismatch problem, we only use sound labels that are related to animal vocalizations and do not overlap with other non-animal sounds. 
After data cleaning, we have 1796 pieces of audio with 37 classes from AudioSet. 
However, even if the sounds have the same label in the AudioSet, they can still sound very different. To alleviate this problem, we treat each audio file in AudioSet as its own class, so we have 1796 classes in this dataset, which is named AudioSet-Aminal-SL, where SL means strongly labeled. To balance the 1796 classes and 47 classes in AudioSet-Aminal-SL and DCASE2022-T5, we choose half classes from each dataset during episodic training.




\subsection{Evaluation metric}
We use the F-measure score, the official evaluation metric provided by the organizers of DCASE task 5, as our main evaluation metric. We also report system performance with the Polyphonic Sound Detection Score~(PSDS)~\cite{bilen2020framework}, which is a robust intersection-based sound event detection evaluation metric. In PSDS, we set the detection tolerance criterion~(DTC) and the ground truth intersection criterion (GTC) to 0.5, and the maximum effective false positive rate to 100.0. Other parameters like the cross-trigger tolerance criterion (CTTC) are not used because our task is not polyphonic detection.

\subsection{Experimental setup}
Following~\cite{yang2021few}, all the audio data are resampled to a 22.5 kHz sampling rate. The input feature of our system is the stack of PCEN~\cite{wang2017trainable} and $\Delta$MFCC~\cite{5709752} features. In the short-time Fourier transform, we set the window length as 1024 and hop size as 256. We set the mel-frequency dimension as 128. 
The input length of our model during training is 0.2 seconds. If the sound event is less than 0.2 seconds, zero-padding will be applied.
The size of the embedding mentioned in Section~\ref{sec:feature-extraction-network} is 2048, in which $C=64,T=4,F=8$.
All the experiments use an initial learning rate of 0.001 with 0.65 exponential decay every 10 epochs. We perform validation after every epoch. We perform validation in a 3-way-5-shot manner since there are only three classes~(HB, ME, PB) in the validation set. We will stop model training if the validation accuracy does not improve for 10 consecutive epochs. And the model with the best validation accuracy is used for evaluation. To make full use of training data, we implement a dynamic data loader that generates training data with a random starting time on the fly. We assume the duration of one vocalization for a certain animal do not vary significantly. Therefore, we design the post-processing strategy for a sound class based on maximum length of positive event $t_{\max}=\max(t_{1},...,t_{K})$. We will remove a positive detection if its length is smaller than $\alpha*t_{\max}$ or greater than $\beta*t_{\max}$. During evaluation, we use $\beta=2.0$, $\alpha=[0.1,0.2,...,0.9]$, and threshold $h=[0.0,0.05,...,0.95]$. We use different combinations of $\beta, \alpha, h$ to calculate data points~\cite{bilen2020framework}, draw the PSD-ROC curve, and calculate PSDS. We choose the best F-measure among all $\beta, \alpha, h$ combinations as the final F-measure. 


\begin{figure}[tbp]
    \centering
    \includegraphics[page=5,width=\columnwidth]{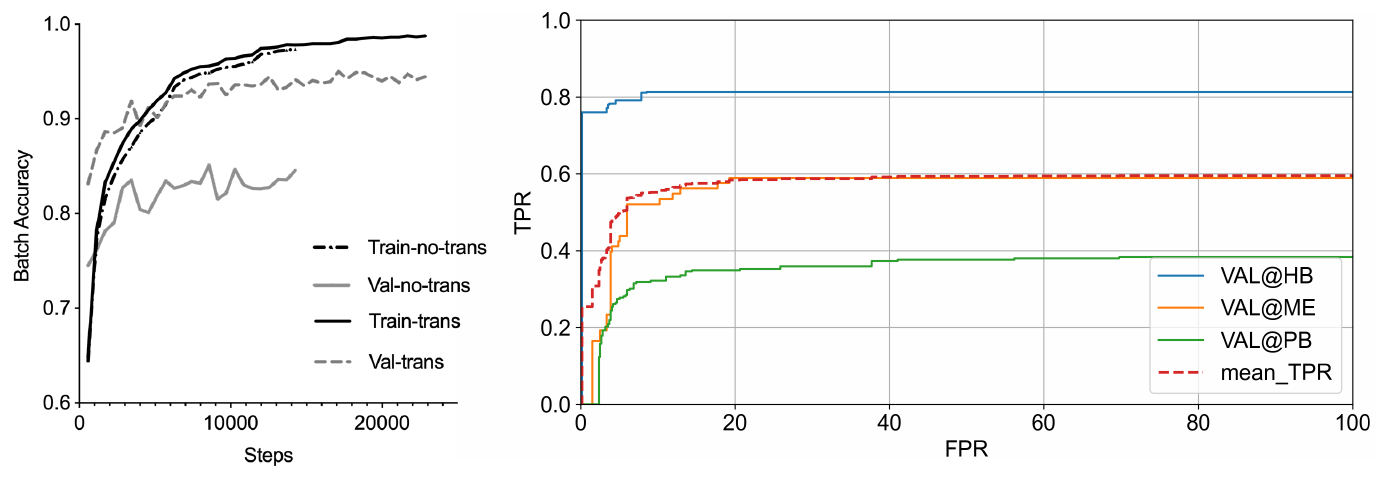}
    \caption{The training and validation accuracy at different training steps, both with and without transductive inference, are shown in the left figure. The right figure is the PSD-ROC curve of our proposed system. \textit{HB}, \textit{ME}, and \textit{PB} are three subsets in the evaluation set. The area under \textit{mean\_TPR} curve is PSDS, which indicates the system overall performance. TPR and FPR stands for true positive rate and false positive rate, respectively. }
    \label{fig:roc}
\end{figure}

\begin{figure}[tbp]
    \centering
    \includegraphics[page=4,width=\columnwidth]{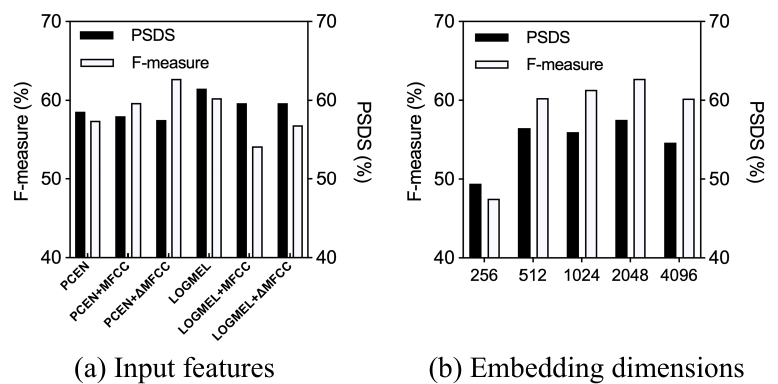}
    \caption{Ablation study on (a)~input features and (b)~embedding dimension. We report the PSDS and F-measure on the DCASE2022-T5 validation dataset.}
    \label{fig:ablation}
\end{figure}

\section{Result}
\label{sec:result}

\begin{table}[htbp]
\begin{tabular}{c|c|c|c|c}
\hline
\multicolumn{1}{c|}{Method} & \multicolumn{1}{c|}{Pre.} & \multicolumn{1}{c|}{Rec. } & \multicolumn{1}{c|}{F-measure } & \multicolumn{1}{c}{PSDS } \\ \hline
Template Matching~\cite{morfi2021few}  & 2.42  & 18.32   & 4.28 & N/A  \\ \hline
ProtoNet (official)~\cite{morfi2021few} & 36.34 & 24.96  & 29.59 & N/A \\ \hline
ProtoNet (our impl) & 23.26 & 63.27  & 34.02 & 46.10 \\ \hline
Proposed         & 69.30    & 57.30     & \textbf{62.73} & \textbf{57.52} \\ \hline
\end{tabular}
\caption{Comparisons with baseline template matching and prototypical network methods. \textit{Pre.} and \textit{Rec.} stand for precision and recall, repectively. The first two methods~\cite{morfi2021few} did not report PSDS result. All the metrics are written in percentage.}
\label{tab:comparison_with_baseline}
\end{table}

The performance of our system on the evaluation set is reported in Table~\ref{tab:comparison_with_baseline}. The F-measure score of template matching and our re-implemented prototypical network baseline~\cite{morfi2021few} is 4.28 and 34.02, respectively. Our system outperforms the baselines by a large margin with an F-measure score of 62.73 and a PSDS of 57.52. 


As is shown in Figure~\ref{fig:roc}, using transductive inference can significantly imrpove the validation accuracy. And the class-wise ROC indicates the \textit{HB} class, which is mostly mosquito sounds, is the easiest one to detect. Class \textit{PB} is the hardest class perhaps because it mainly consists of sparse bird calls with strong background noise. Class \textit{ME} achieves an average performance in the evaluation set.

We perform a study on the effect of the input feature. As shown in Figure~\ref{fig:ablation}(a), the performance of F-measure and PSDS is not always consistent, and we use F-measure to guide our selection considering it is widely used in prior studies~\cite{morfi2021few}. By comparing the F-measure score, PCEN+$\Delta$MFCC appears to be a good feature combination on the evaluation set. We also compare different embedding dimension in Figure~\ref{fig:ablation}(b). We change the dimension by altering the dimension of $F$ in the adaptive average pooling. We notice a dimension of 512 can considerably improve over 256, and 2048 has the best performance among all the settings.




We perform ablations on each of the components we proposed. As shown in Table~\ref{tab:ablation-study}, if we remove the negative segments, the performance drops considerably. The trend is the same with transductive inference and post-processing. We also study the effect of training data. In Table~\ref{tab:data}, we can see that the best F-measure score is achieved using the DCASE2022-T5 only. Using AudioSet-SL leads to an F-measure of 46.83 and a PSDS of 51.00. By combining two datasets we got an F-measure of 58.48 and a best PSDS of 58.77. We hypothesize that the degradation of F-measure using AudioSet is caused by domain mismatch on training data. However, combining two datasets yield the best PSDS, which means using AudioSet data can lead to a general improvement across all threshold and post-processing settings instead of getting a single best system with a high F-measure. This indicates that PSDS might be a suitable metric for the community to reference in this task.

\begin{table}[tbp]
\centering
\begin{tabular}{c|c|c}
\hline
Setting                    & F-measure (\%) & PSDS (\%) \\ \hline
Proposed                   & \textbf{62.73}   & \textbf{57.52}       \\ \hline
 \textbf{w/o} Negative contrast    & 55.25 &  54.95     \\ \hline
 \textbf{w/o} Transductive inference   & 56.37   &  54.50    \\ \hline
 \textbf{w/o} Post processing      & 57.27  &  55.90           \\ \hline
\end{tabular}
\caption{Ablation study of the proposed method.}
\label{tab:ablation-study}
\end{table}

\begin{table}[tbp]
\centering
\begin{tabular}{c|c|c}
\hline
Training data                    & F-measure (\%) & PSDS (\%) \\ \hline
DCASE                   & \textbf{62.73}  &  57.52       \\ \hline
AudioSet-Aminal-SL    & 46.83    & 51.00      \\ \hline
AudioSet-Aminal-SL \& DCASE          & 58.48   & \textbf{58.77}       \\ \hline
\end{tabular}
\caption{A study on using different training datasets. DCASE stands for the DCASE2022-T5 dataset.}
\vspace{-1.0em}
\label{tab:data}
\end{table}



\section{Conclusions}
\label{sec:conclusions}
This paper proposes a new framework for few-shot sound event detection. Our proposed metric learning with negative segments and the transductive inference scheme can significantly improve model performance. On the input feature, our experiment shows that PCEN with $\Delta$MFCC yields the best performance in our settings. Our result also indicates that PSDS might be a useful metric to evaluate the model's overall performance by considering multiple thresholds and post-processing settings during evaluation. 


\newpage

\section{ACKNOWLEDGMENT}
\label{sec:ack}
This research was partly supported by BBC Research and Development, Engineering and Physical Sciences Research Council (EPSRC) Grant EP/T019751/1 "AI for Sound", and a PhD scholarship from the Centre for Vision, Speech and Signal Processing (CVSSP), Faculty of Engineering and Physical Science (FEPS), University of Surrey. For the purpose of open access, the authors have applied a Creative Commons Attribution (CC BY) license to any Author Accepted Manuscript version arising.

\bibliographystyle{IEEEtran}
\bibliography{refs}

%
%
%
%
%
%
%
%
%

\end{sloppy}
\end{document}